\documentclass[aps,prd,nofootinbib,twocolumn]{revtex4-1}
\usepackage{epsfig}
\usepackage[colorlinks,linkcolor=blue,anchorcolor=blue,citecolor=blue,urlcolor=blue,breaklinks=true]{hyperref}
\usepackage{graphics}
\usepackage{slashed}
\usepackage{color}
\usepackage{amsmath}
\usepackage{ulem}
\usepackage{bm}
\usepackage{setspace}
\usepackage{caption}
\usepackage{multirow}
\usepackage{subfigure}

\def\be{\begin{eqnarray}}\def\ee{\end{eqnarray}}

\allowdisplaybreaks

\begin{document}
\author{Cheng-Ming Li$^{1}$}\email{licm@zzu.edu.cn}
\author{Guang-Hao Yu$^{1}$}
\author{Ya-Peng Zhao$^{2}$}\email{zhaoyapeng2013@hotmail.com}
\author{Zhibin Li$^{1}$}\email{lizhibin@zzu.edu.cn}
\author{Jin-Li Zhang$^{3}$}
\author{Yong-Liang Ma$^{4,5}$}
\author{Yong-Feng Huang$^{6,7}$}
\address{$^{1}$ Institute for Astrophysics, School of Physics, Zhengzhou University, Zhengzhou 450001, China}
\address{$^{2}$ School of Mathematics and Physics, Henan University of Urban Construction, Pingdingshan 467036, China}
\address{$^{3}$ School of Mathematics and Physics, Nanjing Institute of Technology, Nanjing 211167, China}
\address{$^{4}$ School of Frontier Sciences, Nanjing University, Suzhou 215163, China}
\address{$^{5}$ International Center for Theoretical Physics Asia-Pacific (ICTP-AP), UCAS, Beijing 100190, China}
\address{$^{6}$ School of Astronomy and Space Science, Nanjing University, Nanjing 210023, China}
\address{$^{7}$ Key Laboratory of Modern Astronomy and Astrophysics (Nanjing University), Ministry of Education, China}
\title{QCD vacuum pressure and its influence on the equation of state of non-strange quark stars}

\begin{abstract}
Solutions of the quark gap equation and the corresponding vacuum pressure are investigated within a modified Nambu-Jona-Lasinio model, which is a basic issue for studying the QCD equation of state (EOS) and the properties of hypothetical non-strange quark stars. In this study,
the coupling strength $G$ is modified as $G=G_1+G_2\langle\bar{\psi}\psi\rangle$ to highlight the feedback effect of the quark condensate on the gluon propagator.
Our analysis reveals that the influence of the vacuum pressure on EOS stiffness critically depends on whether the chiral phase transition is a first-order transition or a smooth crossover. A small ratio $G_1/G$ $(0.74\sim0.75)$ leads to a low vacuum pressure and a first-order chiral phase transition, a scenario favored by the existence of massive pulsars. Conversely, a large $G_1/G$ $(>0.96)$ leads to a high vacuum pressure and a crossover, but the corresponding EOS is ruled out by recent pulsar mass-radius observations. The model parameter space, restricted by four constraints, indicates the current quark mass is in the range $4.08\leq m\leq4.13$ MeV, with the quark condensate feedback contribution accounting for approximately 25\%. Furthermore, it is argued that the merging compact binary in GW170817 could be non-strange quark stars, and the tidal deformability is constrained to $\Lambda(1.4)\leq646$.

\bigskip

\bigskip


\end{abstract}


\maketitle

\section{INTRODUCTION}
The equation of state (EOS) of dense matter plays
an important role in the study of compact stars. Once the EOS is determined, the corresponding
mass-radius and tidal deformability-mass relations can
be obtained by solving the Tolman-Oppenheimer-Volkoff
(TOV) equations coupled with additional differential equations for the tidal Love number. Given that the density is extremely high in the core of compact stars (several times the nuclear saturation density), the deconfined
quark matter dominated by the strong interaction may
exist. Furthermore, there is a possibility that the entire star is composed of pure quark matter, known as
a quark star. Witten and others proposed that strange
quark matter might be the ground state of strongly interacting matter~\cite{10.1143/PTP.44.291,terazawa1979tokyo,PhysRevD.4.1601,Witten1984rs}, while a recent study suggested that stable
quark matter could be non-strange~\cite{PhysRevLett.120.222001}. Due to the lack of
a first-principle understanding of strong interactions at
finite density, the properties of strange compact stars~\cite{PhysRevD.101.063023,GENG2021100152,PhysRevD.108.123031,PhysRevD.111.063033,nj2s-9yy8} and non-strange compact stars~\cite{PhysRevD.103.063018,PhysRevD.106.116009,PhysRevD.103.094037,PhysRevD.108.054013,universe6050063} are usually investigated based on simplified assumptions. Notably, whether two-flavor or three-flavor quark matter is more stable remains an open question till now.

The quantum chromodynamics (QCD) is the fundamental theory of the cold, dense quark matter governed by strong
interactions in compact stars. However, the perturbative expansion is invalid in this regime due to the low energy scale.
Therefore, non-perturbative approaches are necessary to
address this issue. In addition, the ``sign problem''
renders lattice QCD intractable at finite chemical potential and low temperatures. Consequently, in order to
investigate the EOS of QCD in compact stars, we must resort to effective models, such as the chiral effective model~\cite{Brown:2001nh,Holt:2014hma,Ma:2019ery}, quarkyonic model~\cite{McLerran:2007qj,McLerran:2018hbz}, MIT bag model~\cite{HOSAKA199665,PhysRevD.97.083015,Miao_2021,nltd-bc19,q9pg-9ch5},
the Nambu-Jona-Lasinio (NJL) model~\cite{RevModPhys.64.649,Buballa2005205,li2019kurtosis,
PhysRevD.105.094015,Wang:2018sur,g97l-z3yb,zhao2021qcd,zhao2022nonextensive,zhao2023nonextensive}, and the Dyson-Schwinger equations (DSEs)~\cite{ROBERTS1994477,Roberts2000S1,doi:10.1142/S0218301303001326,Roberts200850,Cloet20141}.

It is found that the vacuum pressure (VP) plays a crucial
role in the study of the EOS of QCD. A model independent calculation of the EOS of QCD at finite temperature and chemical potential results in a constant pressure that can be identified as the VP~\cite{doi:10.1142/S0217751X08040457}. At zero temperature and zero chemical
potential, it corresponds to the bag constant in the MIT bag model. The value of VP is model-dependent, and in many previous studies~\cite{PhysRevD.98.083013,PhysRevD.101.063023,Miao_2021,PhysRevD.106.116009,nltd-bc19}, it has been treated as an adjustable parameter, which
has a significant impact on the stiffness of the EOS. Theoretically, the VP describes the pressure difference between the trivial and the non-trivial vacuum,
which corresponds to the Wigner solution ($M_W$) and
Nambu solution ($M_N$) of the quark gap equation, respectively. Based on the property of the Winger solution in the chiral
limit (i.e., the current quark mass $m=0$), it is suggested that in the non-chiral limit (i.e., $m>0$), $M_W$ can be
approximated by the current quark mass $m$~\cite{Buballa2005205}, even though
it is not a rigorous solution to the quark gap equation~\cite{Cui2014,Cui2018,0954-3899-45-10-105001,PhysRevD.99.076006}.

In terms of astronomical observations, since the discovery of the first pulsar in 1967, a vast amount of pulsar mass measurements have been obtained. Notably,
the mass of PSR J0348+0432, as one of the most precise mass measurements to date, reaches $2.01\pm0.04$ solar
mass ($M_{\odot}$)~\cite{Antoniadis1233232}, ruling out a large number of soft EOSs~\cite{doi:10.1146/annurev-astro-081915-023322}. In recent years, the NASA's Neutron Star Interior Composition Explorer (NICER) missions have significantly advanced joint mass-radius observations of pulsars, such as
PSR J0740+6620~\cite{cromartie2020relativistic,Miller_2021,Riley_2021}, PSR J0030+0451~\cite{Miller_2019,Riley_2019}, and HESS J1731-
347~\cite{doroshenko2022strangely}. Specifically, HESS J1731-347, characterized by its
low mass and small radius, imposes stringent constraints
on the EOS of QCD~\cite{su2024quark}. Furthermore, the binary system
merger gravitational wave event GW170817, opened a
new era of multi-messenger astronomy~\cite{PhysRevLett.119.161101}. In this event, the tidal deformability during the inspiral phase of the binary merger was well constrained, ruling out a number of
stiff EOSs~\cite{PhysRevLett.119.161101,PhysRevX.9.011001}.

In this work, we investigate the possible existence of non-strange quark stars by using a modified two-flavor NJL model
that considered the effect of the quark condensate on the effective gluon propagator encoded in the four-fermion interaction coupling constant $G$. Meanwhile, to ensure
thermodynamic consistency, the corresponding thermodynamic potential within the mean-field approximation
is also modified. Under this framework, we solve the
quark gap equation to study the properties of the Winger solution
and Nambu solution and obtain the corresponding VP. For comparison, we analyze the VP under different parameter sets.
Considering the constraints of beta equilibrium and electric charge neutrality in non-strange quark stars, the EOS of non-strange quark matter is derived. Finally, utilizing the observational data on pulsar masses,
radii, and the tidal deformability, we constrain the EOS
and obtain the model parameter space.

The paper is organized as follows. In Sec.~\ref{one}, we briefly introduce the modified NJL model and the Winger solution and Nambu solution
to the quark gap equation. In Sec.~\ref{two}, we present the VP
and the EOS for non-strange quark matter
at zero temperature and finite chemical potential. In
Sec.~\ref{three}, we calculate the mass-radius relation and tidal
deformability of non-strange quark stars, and then constrain the model parameter space using astronomical observations of pulsars.
Finally, a summary and discussion are given in Sec.~\ref{four}.

\section{Solutions to the quark gap equation in the modified NJL model}\label{one}
As a relativistic quantum field theory, the NJL model
is widely used in the study of QCD phase transitions
and compact stars, with the Lagrangian
given by
\be
\label{genLag}
\mathcal{L} =  \bar{\psi}(i{\not\!\partial}-m)\psi
  {} + G \left[(\bar{\psi}\psi)^2+(\bar{\psi}i\gamma^5\tau\psi)^2\right],
\ee
where $m$ is the current quark mass (considering the
isospin symmetry between the $u$ and $d$ quark, $m_u=m_d=m$). In this work, the four-fermion interaction coupling constant $G$
is replaced by
\be
G(\langle\bar{\psi}\psi\rangle) & = & G_1+G_2\langle\bar{\psi}\psi\rangle,
\label{eq:ModifG}
\ee
where $\langle\bar{\psi}\psi\rangle$ refers to the quark condensate.

The modification of the coupling constant $G$ in Eq.~\eqref{eq:ModifG} can be elaborated as follows:
\begin{itemize}
	\item
	
	From the perspective of QCD, the coupling constant $G$ in the NJL model represents an effective gluon propagator. And, by using the coupled DSEs, it is well established that the quark propagator differs significantly between the Nambu phase (characterized by chiral symmetry breaking) and the Wigner phase (characterized by chiral symmetry restoration)~\cite{Cui2018,0954-3899-45-10-105001,PhysRevD.99.076006}. Consequently, the corresponding gluon propagators are expected to differ between these two phases as well~\cite{hong2006quark}. Furthermore, lattice QCD simulations indicate that the gluon propagator exhibits temperature dependence, while its dependence on the chemical potential remains uncertain. In the normal NJL model, the effective gluon propagator $G$ is approximated as a static constant and the difference between Nambu and Wigner phases is not involved.

	\item
	
	From the perspective of the plane wave method in the QCD sum rule~\cite{REINDERS19851}, the full Green function can be decomposed into the perturbative and non-perturbative components, and the condensates are regarded as distinct moments of the non-perturbative Green function.
	Therefore, the most general form of the non-perturbative
	gluon propagator is given by
	\be
	\label{npgp}
	D_{\mu\nu}^{\rm{npert}} & \equiv & D_{\mu\nu}^{\rm{full}}-D_{\mu\nu}^{\rm{pert}}\nonumber\\
	& \equiv &		 a_1\langle\bar{\psi}\psi\rangle+a_2\langle
		G^{\mu\nu}G_{\mu\nu}\rangle+...,
	\ee
	where $\langle G^{\mu\nu}G_{\mu\nu}\rangle$ is the gluon condensate, the coefficients $a_1$ and $a_2$ can be derived with the QCD sum rule approach~\cite{Steele1989,pascual1984qcd}, and the ellipsis refers to contributions from other higher-dimensional condensates. Among all condensates, the quark condensate has the lowest dimension and serves as the order parameter for the chiral phase transition, thus playing a critical role in the QCD sum rule approach. Thus we pick it out, and absorb the effects of all other condensates into the perturbative part of the gluon propagator. This approximation is identical to the modification~\eqref{eq:ModifG}~\cite{PhysRevD.85.034031,Cui2014,PhysRevD.93.036006,PhysRevD.97.103013,Li_2025}.

\end{itemize}

Under the modification~\eqref{eq:ModifG}, the coupling strength $G$ becomes dependent on the quark condensate. Specifically,
$G_2\langle\bar{\psi}\psi\rangle/G$ corresponds to the fraction of the contribution
from the quark condensate to the effective gluon propagator, while $G_1/G$ represents the fraction from other condensates, thus there is a constraint between these two
ratios, i.e., $G_2\langle\bar{\psi}\psi\rangle/G+G_1/G=1$. For convenience, we
adopt the easily assignable ratio $G_1/G$ as an independent parameter, which allows us to constrain its value
via pulsar observations and subsequently limit the
magnitude of $G_2\langle\bar{\psi}\psi\rangle/G$.

To ensure thermodynamic consistency, under this
modification scheme, we also need to apply a corresponding treatment to the term $G\langle\bar{\psi}\psi\rangle^2$
in the mean-field thermodynamic potential of the normal NJL model. Inspired
by a previous work~\cite{PhysRevD.52.5206}, the term $G\langle\bar{\psi}\psi\rangle^2$ can be replaced by
an integral of $2\langle\bar{\psi}\psi\rangle{\rm d}(G\langle\bar{\psi}\psi\rangle)/{\rm d}\langle\bar{\psi}\psi\rangle$ over $\langle\bar{\psi}\psi\rangle$. In this way, the mean-field thermodynamic potential can be derived as
\begin{eqnarray}
  \Omega(T,\{\mu\},\{\langle\bar{\psi}\psi\rangle\}) & = &
  \Omega_{\rm M}(T,\mu)
    +G_1\langle\bar{\psi}\psi\rangle^2
    +\frac{4}{3}G_2\langle\bar{\psi}\psi\rangle^3
  \nonumber\\
  & &{} + {\rm const.},\label{thermpot}
\end{eqnarray}
where $\Omega_{\rm M}$ represents the free Fermi-gas contribution,
\begin{eqnarray}
  \Omega_{\rm M} & =&{} -2N_{\rm c}N_{\rm f}\int\frac{{\rm d}^3
  p}{(2\pi)^3}\{T\,{\rm ln}[1+{\rm
  e}^{-(E_{p}-\mu)/T}]\nonumber\\
  &&{} +T\,{\rm ln}[1+{\rm e}^{-(E_{p}+
  \mu)/T}]+E_{p}\}.\label{thermopotofquasi}
\end{eqnarray}
Here $E_{p}=\sqrt{\overrightarrow{p}^2+M^2}$ is the quark on-shell energy and $M$ refers to the effective quark mass. Notably, when $G_1=G$ and $G_2=0$, the modified version reproduces the
normal NJL model.

Within the mean-field approximation, the quark gap
equation is given by
\begin{eqnarray}
  F(M) &=& m-2
    (G_1+G_2\langle\bar{\psi}\psi\rangle)
  \langle\bar\psi\psi\rangle-M=0.\label{gapeq}
\end{eqnarray}
The quark condensate $\langle\bar{\psi}\psi\rangle$ and the total quark number density $\rho$ are given by
\begin{eqnarray}
  \langle\bar\psi\psi\rangle
       &=&{} -\frac{N_{\rm c}N_{\rm f}M}{\pi^2}
        \int_0^{\pi}{\rm d}p [1-n_{p}(T,\mu)-\overline{n}_{p}(T,\mu)]\frac{p^2}{E_p},\nonumber\\
        \label{psibarpsi} \\
  \rho
       &=& \frac{N_{\rm c}N_{\rm f}}{\pi^2}
        \int_0^{\pi}{\rm d}p
           (n_{p}(T,\mu)-\overline{n}_{p}(T,\mu))p^2,
        \label{rho}
\end{eqnarray}
where $n_{p}(T,\mu)$ and $\overline{n}_{p}(T,\mu)$ are the Fermi occupation
numbers of quarks and antiquarks, respectively, which
are defined as
\begin{eqnarray}
  n_{p}(T,\mu) &=&
  [{\rm e}^{(E_{p}-\mu)/T}+1]^{-1},\label{fonq} \\
  \overline{n}_{p}(T,\mu) &=&
  [{\rm e}^{(E_{p}+\mu)/T}+1]^{-1}.\label{fonaq}
\end{eqnarray}

To investigate the vacuum properties at $T=\mu=0$
within the modified NJL model, we need to solve Eq.~(\ref{gapeq}). Since the NJL model is non-renormalizable, a regularization scheme is required to avoid ultraviolet divergences. In this work, we employ
the three-momentum cutoff regularization and introduce
the ultraviolet momentum cutoff $\Lambda_{\rm UV}$. The model parameters, including $m$, $\Lambda_{\rm UV}$, and $G$, should be fitted to
reproduce the pion mass and decay constant ($M_{\pi}=135$ MeV, $f_{\pi}=92.4$ MeV), where the parameter $m$ is fixed
before the fitting. It is worth noting that the Review of Particle Physics constrained the current quark
mass as $m=3.5^{+0.7}_{-0.2}$ MeV~\cite{olive2014review}. In this work, we take the current quark mass as $m=4.1$ MeV, and the corresponding parameter set is $\Lambda_{\rm UV}=758$ MeV, $G=3.31$ GeV$^{-2}$, $-\langle\bar{u}u\rangle^{1/3}=267$ MeV.

\begin{figure}
\includegraphics[width=0.47\textwidth]{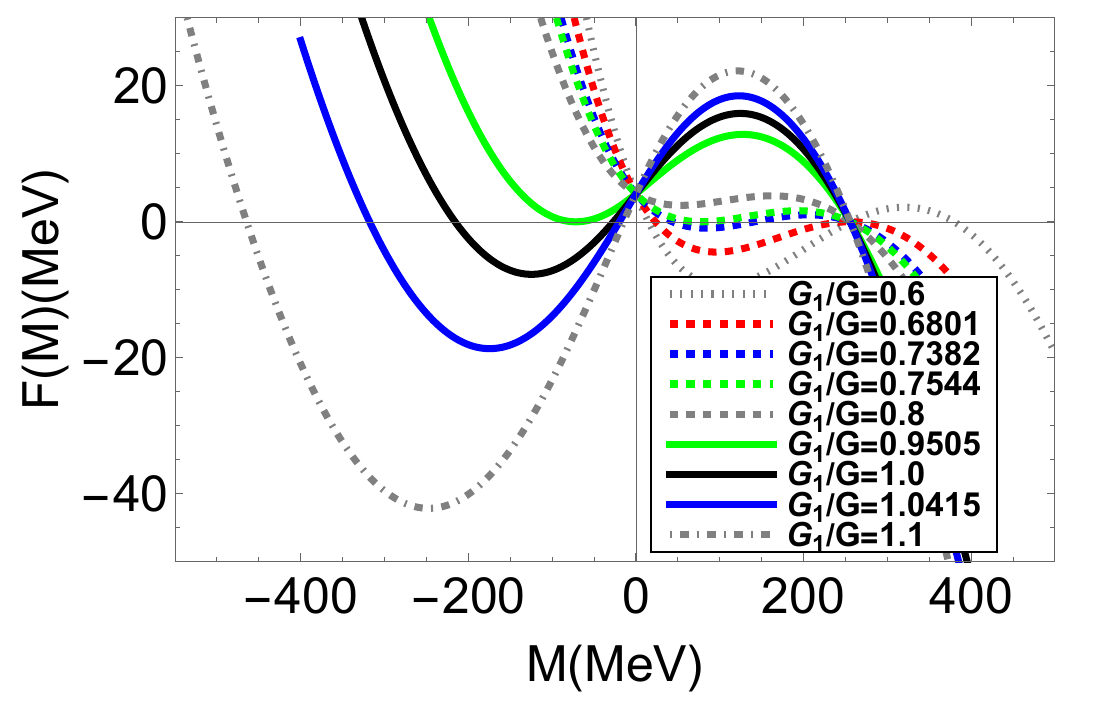}
\caption{ The variation of solutions to Eq.~(\ref{gapeq}) with different $G_1/G$ at $T=\mu=0$.}
\label{Fig:1}
\end{figure}

\begin{figure}
\includegraphics[width=0.47\textwidth]{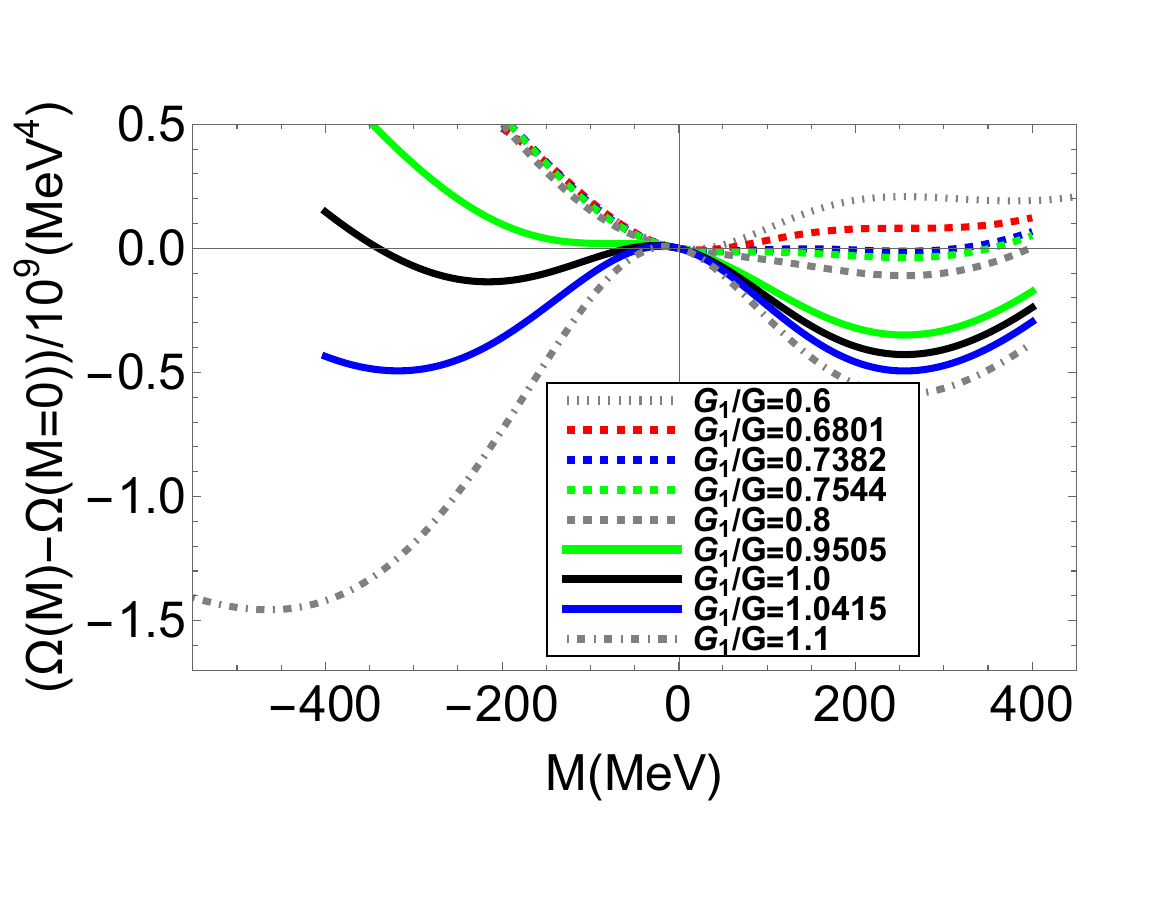}
\caption{The thermodynamic potential as a function of
the effective quark mass with different $G_1/G$ at $T=\mu=0$.}
\label{Fig:2}
\end{figure}

To study the influence of the ratio $G_1/G$ on the solutions to Eq.~(\ref{gapeq}),
we vary it in the range of $0.6\sim1.1$ and present the corresponding $F(M)-M$ relations in Fig.~\ref{Fig:1}. Furthermore, to identify the locations of the Nambu solution (the stable physical solution corresponding to chiral symmetry breaking) and the Winger solution (the physical solution corresponding to chiral symmetry restoration or partial restoration) of Eq.~(\ref{gapeq}), we present the $(\Omega(M)-\Omega(M=0))-M$ relations in
Fig.~\ref{Fig:2}.

In Fig.~\ref{Fig:1}, the intersections of the $F(M)$ curves with the $M-$axis for different $G_1/G$ parameters correspond to
the solutions of Eq.~(\ref{gapeq}). It can be seen that all curves intersect the $M-$axis at $M=255.6$ MeV. This point corresponds to the physical solution of the effective quark mass obtained from fitting the experimental pion data, thus the stable Nambu solution should be located here. When $G_1/G=0.6$, as seen in Fig.~\ref{Fig:2}, $M=255.6$ MeV corresponds to a local maximum of the thermodynamic potential, making it an unstable solution that does not satisfy
the requirements. For $0.6801\leq G_1/G\leq0.7382$, the
Nambu solution appears at $M=255.6$ MeV, but its thermodynamic potential remains higher than that of the Winger solution on the
left, thus still corresponding to an unstable solution. For $0.7382\leq G_1/G\leq0.7544$ and $0.9505\leq G_1/G\leq 1.0415$,
stable Nambu solution appears at $M=255.6$ MeV and the Winger solution
exists. The difference is that $M_W>0$ in the former, while $M_W<0$ in the latter. To compare
the effects of the variation in $G_1/G$ across these two
ranges, we take five parameter sets corresponding to
$G_1/G=0.74, 0.75, 0.96, 0.97, 1$ for the subsequent calculations.

\section{The QCD vacuum pressure and EOS of non-strange quark matter}\label{two}
\begin{figure}
\centering
\subfigure[]{
\centering
\label{Fig:3a}
\includegraphics[width=0.45\textwidth]{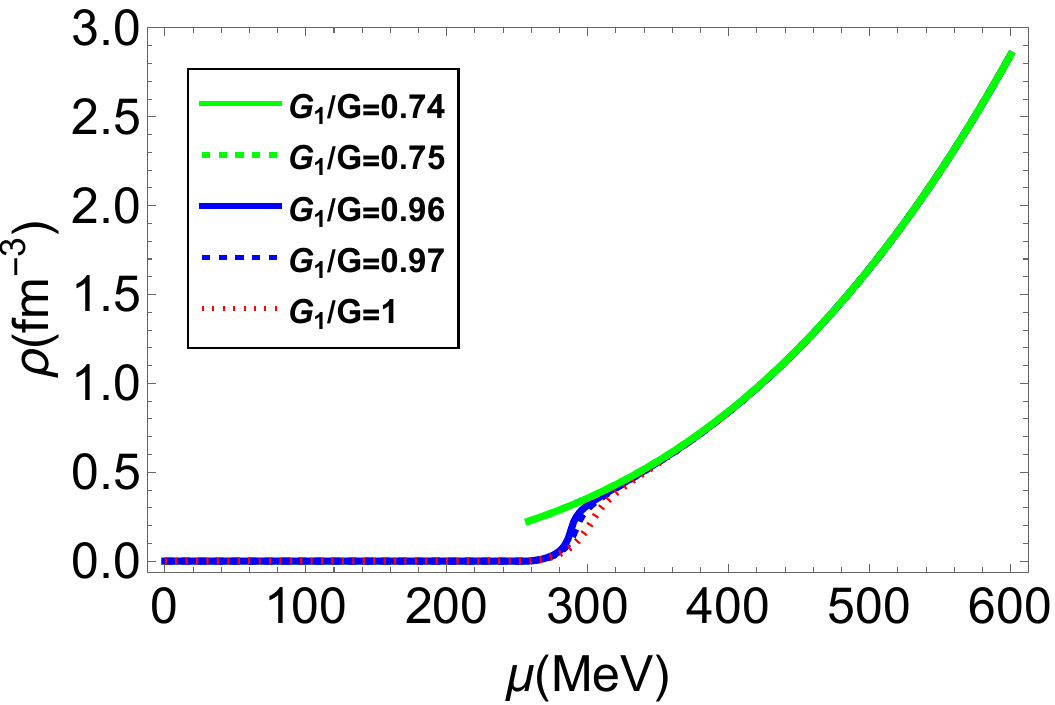}
}
\subfigure[]{
\centering
\label{Fig:3b}
\includegraphics[width=0.45\textwidth]{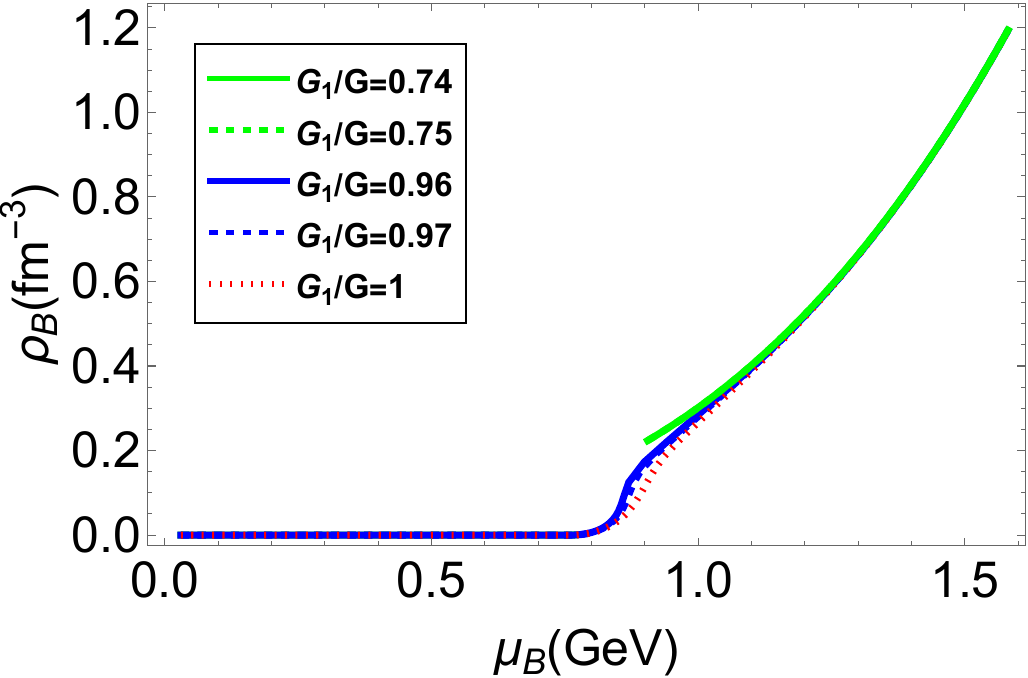}
}
\caption{Number density at $T = 0$: (a) the quark number density without $\beta-$equilibrium and charge neutrality, (b) the baryon number density with $\beta-$equilibrium and charge neutrality.}
\label{Fig:3}
\end{figure}

For non-strange quark stars, the temperature is negligible compared to the high chemical potential, allowing
them to be well described by the zero-temperature approximation. At $T = 0$, we calculate the relationship
between the particle number density and the chemical
potential of quark matter, and the results are shown
in Fig.~\ref{Fig:3}. In Fig.~\ref{Fig:3a}, we find that the curves for $G_1/G = 0.74, 0.75$ almost coincide, and there are discontinuous jumps as the chemical potential increases, corresponding to the occurrence of first-order chiral phase transitions. In contrast, the curves for $G_1/G = 0.96, 0.97, 1$ show that the quark number densities increase smoothly as the chemical potential increases, corresponding to smooth crossovers. The critical chemical potentials (corresponding to the discontinuous or inflection points of the
curves) increase as $G_1/G$ increases.

In quark stars, the $\beta-$equilibrium and electric charge neutrality should be taken into account, which is
\begin{eqnarray}\label{constrains}
  &&\mu_{\rm d}=\mu_{\rm u}+\mu_{\rm e}, \nonumber \\
  &&\frac{2}{3}\rho_{\rm u}-\frac{1}{3}\rho_{\rm
  d}-\rho_{\rm e}=0,
\end{eqnarray}
where $\rho_{\rm e}=\mu_{\rm e}^3/3\pi^2$ represents the number density of
electrons at $T = 0$. The relationship between the baryon number density and the baryon chemical potential of
quark matter is shown in Fig.~\ref{Fig:3b}, where $\mu_{\rm B}=\mu_{\rm u}+2\mu_{\rm d}$, $\rho_{\rm B}=(\rho_{\rm u}+\rho_{\rm d})/3$. One can see that, for $G_1/G = 0.74$ and $0.75$, there are gaps within the interval $768<\mu_B<905$ MeV as the first-order chiral phase transition occurs, while for $G_1/G = 0.96, 0.97$ and $1$, a smooth crossover occurs and there are no gaps.

At finite temperature and finite chemical potential, the
pressure can be expressed as~\cite{doi:10.1142/S0217751X08040457}
\begin{equation}\label{eos}
  P(\mu, T)=P(\mu=0, T)+\int_0^{\mu}{\rm d}\mu^\prime\rho(\mu^\prime, T),
\end{equation}
which is the so-called EOS of QCD. It can be seen that
the pressure consists of two parts: the first term is the constant pressure at $\mu = 0$ that corresponds to the negative VP, and the second term is the non-trivial, $\mu-$dependent part. In the phenomenological MIT bag model~\cite{HOSAKA199665}, free quarks are considered to be confined within ``bags'' of non-trivial vacuum. The pressure difference between these bags and the external trivial vacuum corresponds to the VP of QCD, which is also called the bag constant $-B$. Theoretically, the trivial vacuum and non-trivial vacuum can be described by the Winger solution and Nambu solution, respectively. Following the suggestion in Refs.~\cite{0954-3899-45-10-105001,Cui2018,PhysRevD.99.076006}, the VP is defined as
\begin{eqnarray}
\label{Bagconstant}
P(\mu=0, T) & = & P(M_W)-P(M_N) \nonumber\\
            & = &\Omega(\mu=0,T,\langle\bar{\psi}\psi\rangle_N) \nonumber\\
            & &{} - \Omega(\mu=0,T,\langle\bar{\psi}\psi\rangle_W),
\end{eqnarray}
where $M_W$ and $M_N$ are the Winger solution and Nambu solution
of Eq.~(\ref{gapeq}), and $\langle\bar{\psi}\psi\rangle_W$ and $\langle\bar{\psi}\psi\rangle_N$ are the corresponding quark condensates.

\begin{figure}
\centering
\subfigure[]{
\centering
\label{Fig:4a}
\includegraphics[width=0.45\textwidth]{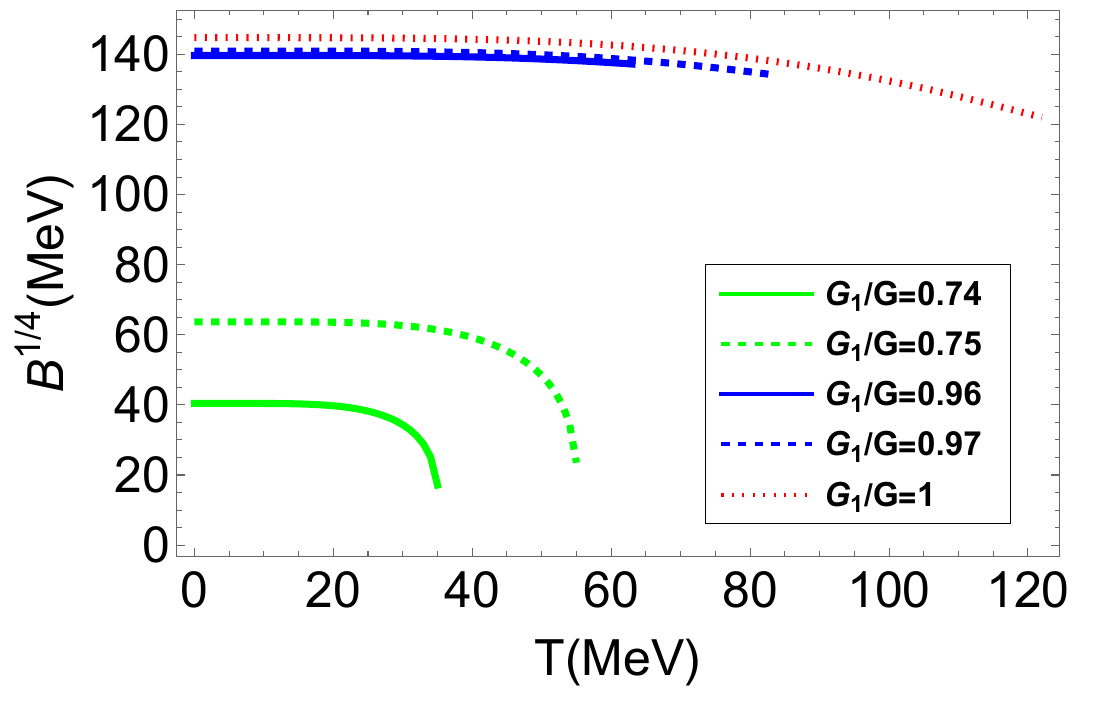}
}
\subfigure[]{
\centering
\label{Fig:4b}
\includegraphics[width=0.45\textwidth]{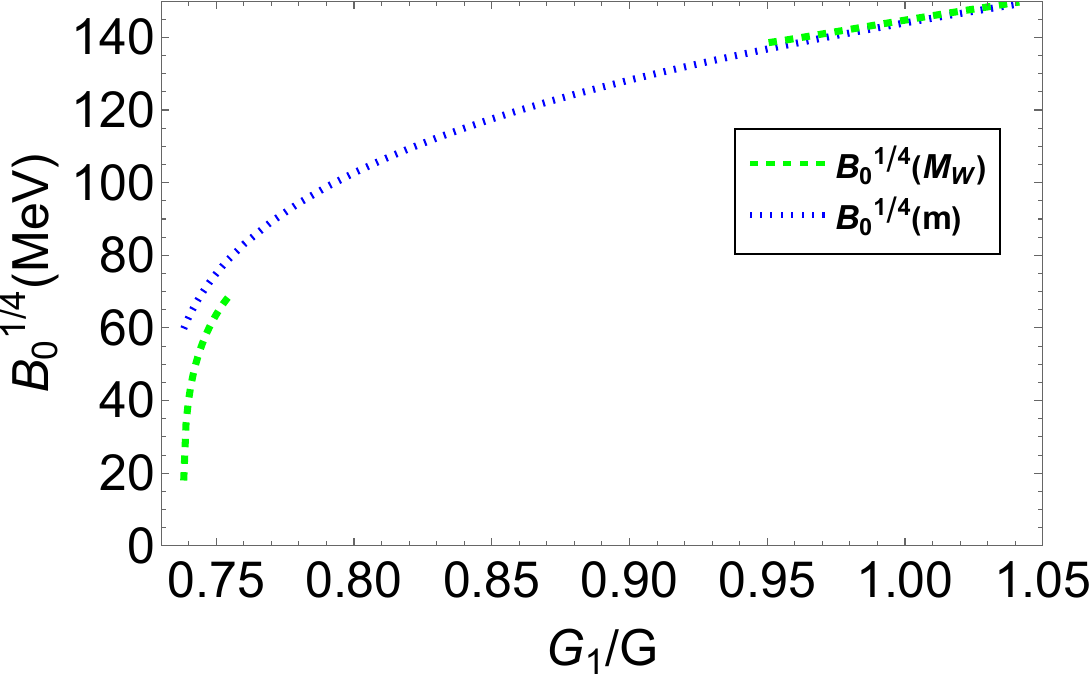}
}
\caption{(a) $B^{1/4}-T$ relations for different $G_1/G$, and (b) $B_0^{1/4}-G_1/G$ relations using the strict Winger solution and the approximate Winger solution, denoted as $B_0^{1/4}(M_W)$ and $B_0^{1/4}(m)$, respectively.}
\label{Fig:4}
\end{figure}

Since $B^{1/4}$ has also been commonly referred to as the bag constant in previous studies, for the sake of distinction, we will use $B$ to denote the magnitude of the VP and $B^{1/4}$ to specifically refer to the bag constant in the following. In Fig.~\ref{Fig:4}, we present the calculation results for $B^{1/4}$. In Fig.~\ref{Fig:4a},
it can be found that $B^{1/4}$ gradually decreases with increasing temperature, where variation for the curves with $G_1/G = 0.74, 0.75$ is more pronounced than that for $G_1/G = 0.96, 0.97, 1$. Furthermore, the difference between the former two curves is significantly larger than that among the latter three.

Considering the zero-temperature approximation in non-strange quark stars, we focus primarily on the results of the EOS of QCD
at $T = 0$. Accordingly, we investigate the bag constant
at $T = 0$, as shown in Fig.~\ref{Fig:4b}. It is known that in
the chiral limit ($m = 0$), $\langle\bar{\psi}\psi\rangle=0$ implies chiral symmetry restoration, and the Winger solution in this case is $M_W = m = 0$. Therefore, it is suggested that in the non-chiral limit ($m>0$), $M_W = m$ can be used as an approximation for the Winger solution~\cite{Buballa2005205} (although it is not the strict Winger solution in this case).

In Fig.~\ref{Fig:4b}, we compare the calculation results of $B^{1/4}$ at $T = 0$ using the strict and the approximate Winger solutions, denoted as $B_0^{1/4}(M_W)$ and $B_0^{1/4}(m)$, respectively. For
$0.74\lesssim G_1/G\lesssim0.75$, $B_0^{1/4}(M_W)<B_0^{1/4}(m)$, whereas for $G_1/G\gtrsim0.96$, $B_0^{1/4}(M_W)>B_0^{1/4}(m)$.
In addition, the discrepancy between the two $B_0^{1/4}$, i.e., $|B_0^{1/4}(M_W)-B_0^{1/4}(m)|$, is significantly larger for a small $G_1/G$ than that for a large $G_1/G$. In particular, for $G_1/G=1$, $|B_0^{1/4}(M_W)-B_0^{1/4}(m)|$ is very small, reflecting the validity of adopting the approximate Winger solution in the normal NJL model. Similar to the results in Refs.~\cite{0954-3899-45-10-105001,PhysRevD.99.076006}, under the non-chiral limit ($m>0$), we find that the Winger solution also disappears at high temperatures. Consequently, discontinuities appear in Fig.~\ref{Fig:4}, indicating that $B^{1/4}$
cannot be calculated numerically in this regime.

\begin{figure}
\includegraphics[width=0.47\textwidth]{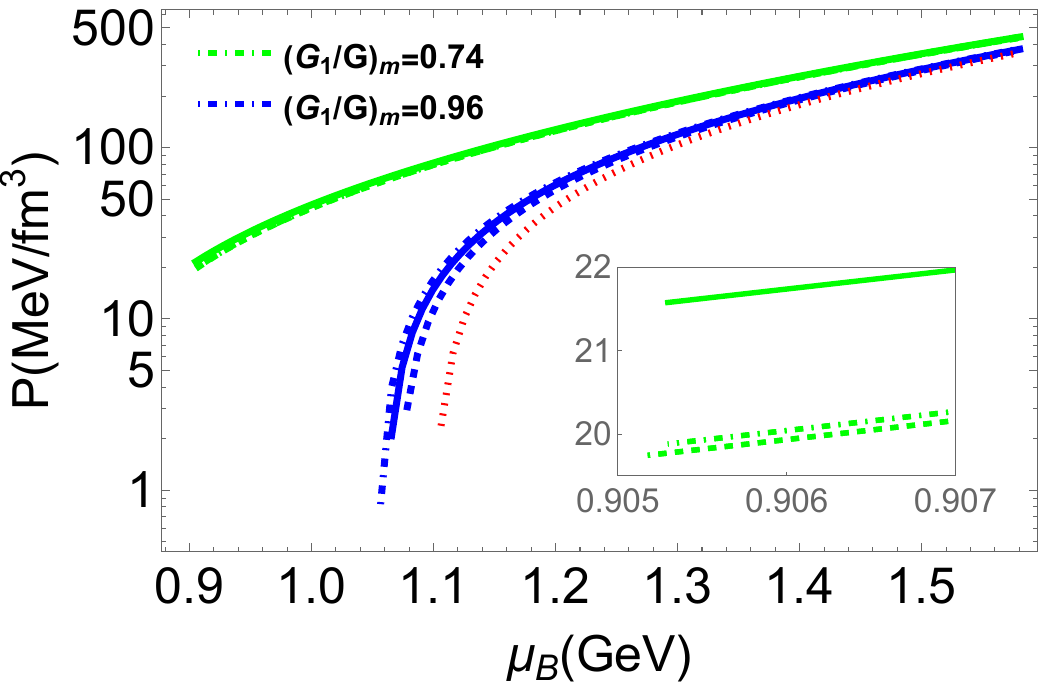}
\caption{The EOS of QCD. The green and blue
dash-dotted line represents the calculations using the
approximate Winger solution with $G_1/G = 0.74, 0.96$,
respectively. The other five lines have the same
meanings as those in Fig.~\ref{Fig:4a}, that is, the calculations using the strict Winger solution with
$G_1/G = 0.74, 0.75, 0.96, 0.97, 1$.}
\label{Fig:5}
\end{figure}

In Fig.~\ref{Fig:5}, we present the results for the EOS. We can see that for the same baryon chemical potential, a larger $G_1/G$ leads to a lower pressure. Furthermore, as previously mentioned, the total pressure consists of two contributions: the VP and the non-trivial density-dependent term. It can be inferred from Fig.~\ref{Fig:4b} and Fig.~\ref{Fig:3b} that $G_1/G$ affects both contributions. To highlight the influence of
VP on the EOS, we also display in Fig.~\ref{Fig:5} the results calculated with different bag constants. Since the
VP exists as the subtrahend ($-B$) in the EOS, corresponding to a negative contribution, we have $P(B_0(M_W))>P(B_0(m))$ for $G_1/G=0.74$, and $P(B_0(M_W))<P(B_0(m))$ for $G_1/G=0.96$.

\begin{figure}
\includegraphics[width=0.47\textwidth]{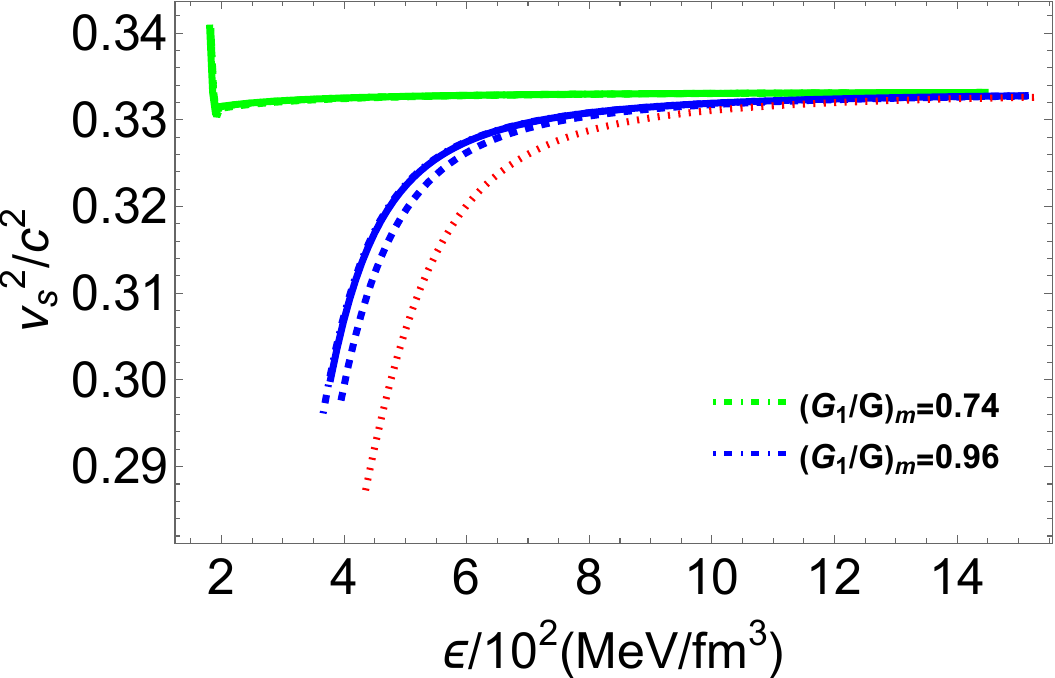}
\caption{The squared sound velocities of the EOSs
shown in Fig.~\ref{Fig:5}.}
\label{Fig:6}
\end{figure}

According to thermodynamic relations, the energy
density of the quark system is given by~\cite{PhysRevD.86.114028,PhysRevD.51.1989}
\begin{equation}\label{rbedasp}
  \epsilon=-P+\sum_{i=u,d,e}\mu_{\rm i}\rho_{\rm i}.
\end{equation}
The sound velocity, which reflects the stiffness of the EOS, is defined as
\begin{equation}\label{soundvelocity}
 \nu_{\rm s} = \sqrt{\frac{{\rm d}p}{{\rm d}\epsilon}}.
\end{equation}
In Fig.~\ref{Fig:6}, we present the $\nu_s^2/c^2-\epsilon$ relation for the
quark system. It can be found that for different values of $G_1/G$, the sound velocities do not exceed
the speed of light and approaches the conformal limit, $\nu_s^2/c^2=1/3$, from below as energy density increases. For example, when $\epsilon\gtrsim1050$ MeV/fm$^3$, $\nu_s^2/c^2\gtrsim0.33$.
For $G_1/G\geq0.96$, a smooth crossover occurs, and the sound velocities decrease as $G_1/G$ increases. For $0.74\leq G_1/G\leq0.75$, the sound velocities exhibit kinks when $\epsilon\sim200$ MeV/fm$^3$, which correspond to the discontinuities in Fig.~\ref{Fig:3} where the first-order chiral phase transition occurs. It is worth noting that for small $G_1/G$, the variation in $G_1/G$ has a negligible influence on the results. Similarly, employing different calculation methods for the bag constant, $B_0^{1/4}(M_W)$ and $B_0^{1/4}(m)$, has a minimal effect on the results. More precise numerical calculations indicate that when $G_1/G$ is large, we have $\nu_s^2(B_0^{1/4}(M_W))<\nu_s^2(B_0^{1/4}(m))$. However, when $G_1/G$ is small, we have $\nu_s^2(B_0^{1/4}(M_W))<\nu_s^2(B_0^{1/4}(m))$ before the first-order phase transition, whereas $\nu_s^2(B_0^{1/4}(M_W))>\nu_s^2(B_0^{1/4}(m))$ after the phase transition. This implies that a larger bag constant leads to a stiffer EOS before the first-order phase transition, but a softer EOS after the phase transition. Similarly, the results for the sound velocities in the cases of $G_1/G = 0.74, 0.75$ show that a larger $G_1/G$ also makes the EOS stiffer before the first-order phase transition, while softer after the phase transition.

In fact, through theoretical analysis, we can demonstrate the influence of the bag constant on the sound velocity described above. Theoretically, merely
changing the magnitude of the bag constant does not affect the $\rho_B -\mu_B$ relation of the system. According to thermodynamic relations, we have $\nu_{\rm s}^2 = (1/\mu_B)\cdot{\rm d}P/{\rm d}\rho_B$. For a fixed $\mu_B$, changing the bag constant corresponds to
altering a constant term in the pressure, thus not affecting the magnitude of sound velocity. However, by combining Eq.~(\ref{eos}) with Eq.~(\ref{rbedasp}) at $T = 0$, we obtain
\be
\epsilon=B(T=0)+\sum_{i=u,d,e}(\mu_{\rm i}\rho_{\rm i}-\int_0^{\mu_{\rm i}}\rho(\mu_{\rm i}^\prime)d\mu_{\rm i}^\prime).
\ee
If the bag constant increases, the energy density will increase accordingly. As a reslut, increasing the bag constant causes a rightward shift of the $\nu_s^2/c^2-\epsilon$ curve. Considering the
characteristics of the $\nu_s^2/c^2-\epsilon$ curves during the smooth crossover and the first-order phase transition in Fig.~\ref{Fig:6},
the conclusions regarding the influence of the bag constant on the sound velocity can be derived naturally.

\section{Structure of non-strange quark stars}\label{three}
\begin{figure}
\centering
\subfigure[]{
\centering
\label{Fig:7a}
\includegraphics[width=0.45\textwidth]{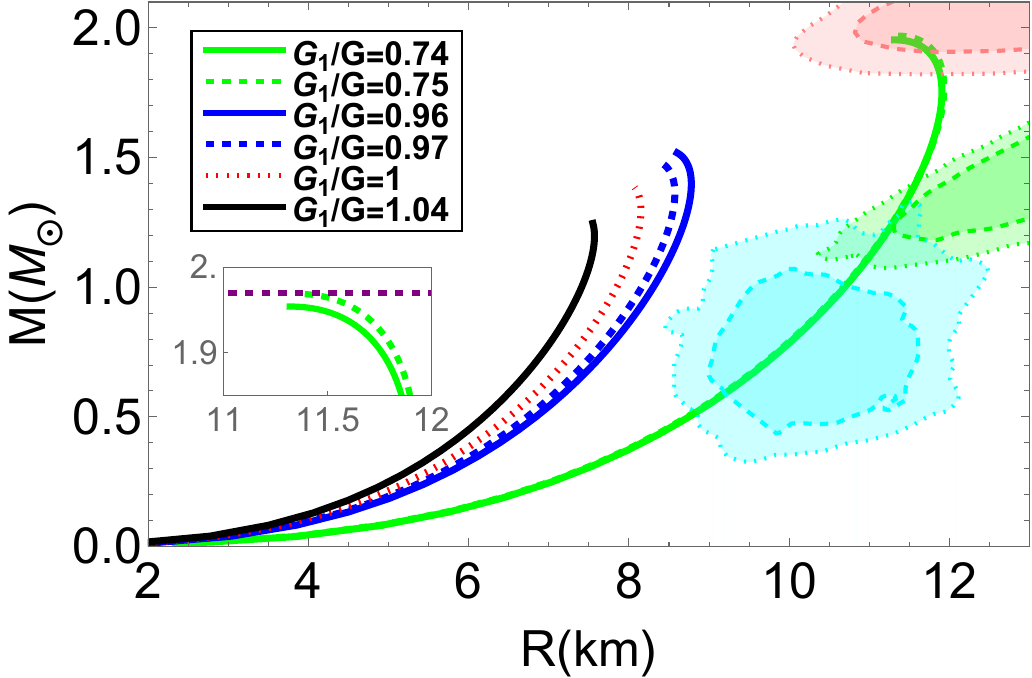}
}
\subfigure[]{
\centering
\label{Fig:7b}
\includegraphics[width=0.45\textwidth]{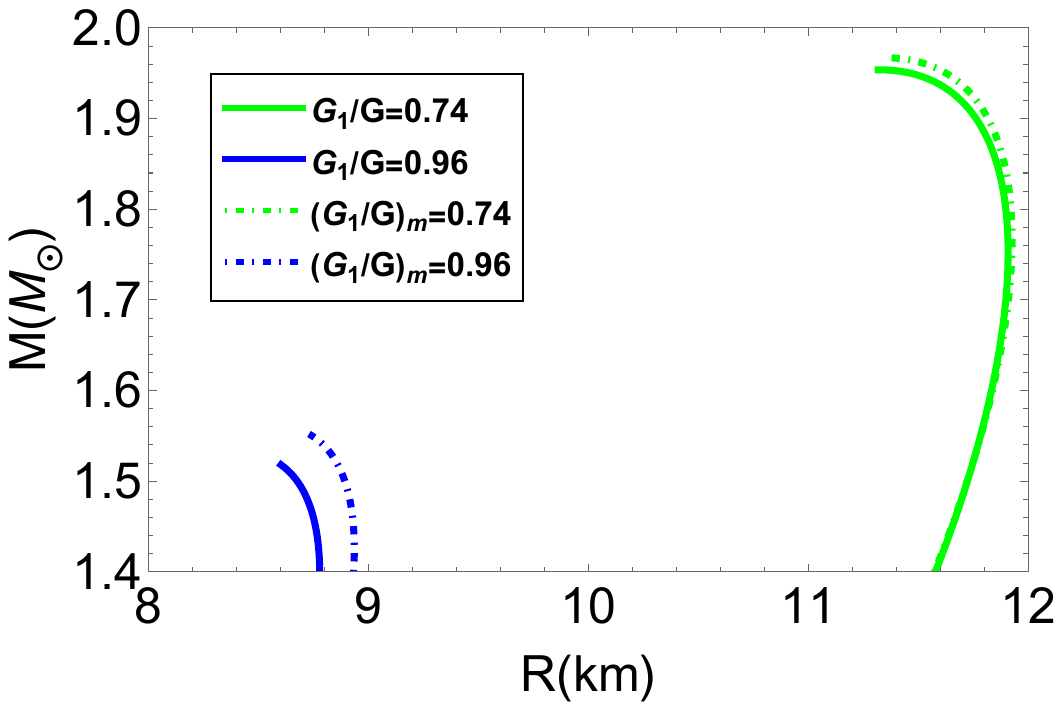}
}
\subfigure[]{
\centering
\label{Fig:7c}
\includegraphics[width=0.45\textwidth]{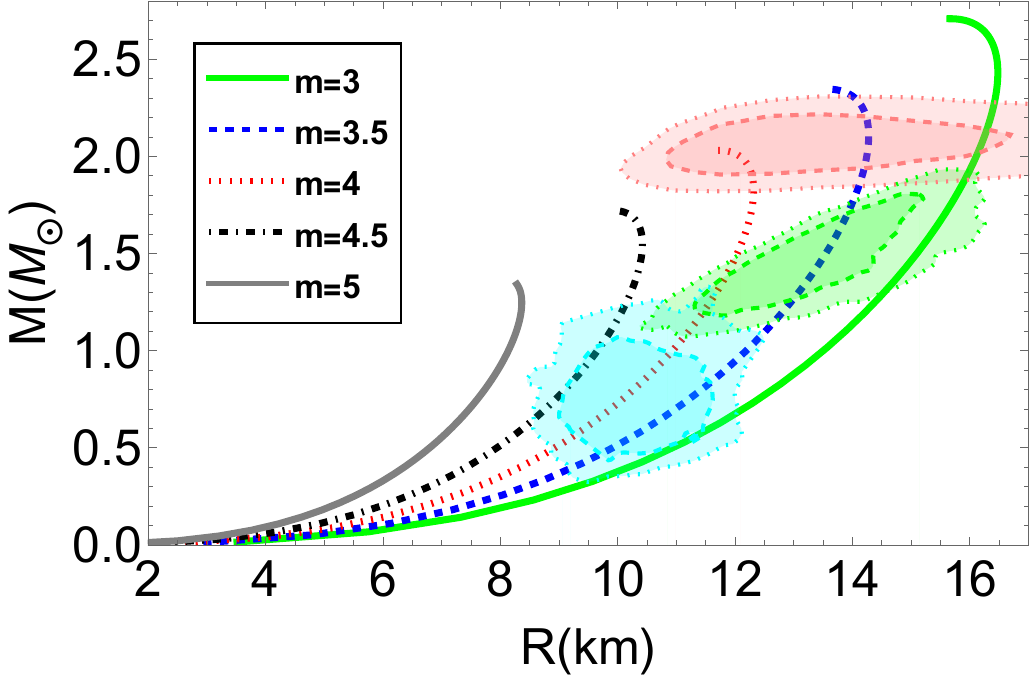}
}
\caption{The $M-R$ relations of non-strange quark
stars, (a) for different $G_1/G$ with the current quark mass $m = 4.1$ MeV, (b) for $G_1/G=0.74$ (green line) and $G_1/G=0.96$ (blue line) using the strict Wigner solution (solid line) and approximate Wigner solution (dash-dotted line), respectively,
(c) for different
current quark mass $m$ whose corresponding maximum mass of the star is the most massive one. The red, green, and blue regions correspond to the constraints from NICER
missions of PSR J0740+6620~\cite{Miller_2021}, PSR J0030+0451~\cite{Miller_2019}, and HESS J1731-347~\cite{doroshenko2022strangely}, respectively.}
\label{Fig:7}
\end{figure}

To investigate the mass-radius relation, we need to
solve the TOV equations,
\begin{eqnarray}
  \frac{{\rm d}P(r)}{{\rm d}r} & = &{} -\frac{G(\epsilon+P)(M+4\pi r^3P)}{r(r-2GM)} \,\, ,\nonumber\\
  \frac{{\rm d}M(r)}{{\rm d}r} & = & 4\pi r^2\epsilon\,\,\, .\label{TOV}
\end{eqnarray}
The results are shown in Fig.~\ref{Fig:7}.

In Fig.~\ref{Fig:7a}, we can see the influence of the $G_1/G$ value on the mass-radius relation. For $0.74\leq G_1/G\leq0.75$, a larger $G_1/G$ leads to a larger maximum mass, whereas for $G_1/G\geq0.96$, a larger $G_1/G$ results in a smaller maximum mass. Notably, for $G_1/G = 0.75$, the corresponding result simultaneously satisfy the constraints from NICER missions of PSR J0740+6620, PSR J0030+0451, HESS J1731-347, and from the mass measurement of PSR J0348+0432. In Fig.~\ref{Fig:7b}, to highlight the influence of VP
on the mass-radius relation, we focus on the differences arising from the bag constant with different calculation methods, $B_0^{1/4}(M_W)$ and $B_0^{1/4}(m)$. For small $G_1/G$ ($=0.74$), a larger VP with $B_0^{1/4}(m)=62.9$ MeV leads to a larger maximum mass, but for large $G_1/G$ ($=0.96$), the reverse is true, i.e., a larger VP
with $B_0^{1/4}(M_W)=139.7$ MeV leads to a smaller maximum mass. Furthermore, the difference in maximum mass caused by the variation in the bag constant increases as $G_1/G$ increases. In Fig.~\ref{Fig:7c}, we investigate the effect of different current quark masses on the mass-radius relation. Each curve in Fig.~\ref{Fig:7c} corresponds to the scenario where the maximum mass of non-strange quark stars reaches its peak for that specific current quark mass. It can be seen that the maximum mass decreases as the current quark mass increases. Only for $m < 4.5$ MeV is it possible to satisfy the mass constraint from PSR J0348+0432~\cite{Antoniadis1233232}, and the constraints from NICER missions of PSR J0740+6620~\cite{Miller_2021}, PSR J0030+0451~\cite{Miller_2019}, and HESS J1731-347~\cite{doroshenko2022strangely}. For $m = 4$ MeV, the maximum mass of non-strange quark stars is $2.03 M_{\odot}$.

In the binary compact star merger gravitational wave
event GW170817, the tidal deformability $\Lambda$ of the binary system during the inspiral phase has been constrained~\cite{PhysRevLett.119.161101,PhysRevX.9.011001}. Theoretically, it is related to the dimensionless tidal Love number $k_2$ for $l = 2$ by the following relation~\cite{PhysRevLett.119.161101},
\begin{equation}\label{td}
 \Lambda=\frac{2}{3} k_2\left(\frac{R}{GM}\right)^5.
\end{equation}
According to Ref.~\cite{PhysRevD.81.123016}, $k_2$ can be calculated by
\begin{eqnarray}
  &k_2&=\frac{8C^5}{5}(1-2C)^2[2+2C(y-1)-y]\nonumber\\
  & & {} \times\{2C[6-3y+3C(5y-8)]\nonumber\\
  & &{}\qquad + 4C^3[13-11y+C(3y-2)+2C^2(1+y)]\nonumber\\
  & &{}\qquad +3(1-2C)^2[2+2C(y-1)-y]ln(1-2C)\}^{-1},\label{tln}\nonumber\\
\end{eqnarray}
where $C = M/R$ is the compactness of the star, and
$y$ for quark stars is defined as~\cite{PhysRevD.81.123016}
\begin{equation}\label{parametery}
  y=R\beta(R)/H(R)-4\pi R^3\epsilon_0/M.
\end{equation}
Here $\epsilon_0$ is the energy density inside the star surface. The values of the metric functions at the surface, $H(R)$ and $\beta(R)$, can be obtained by solving the following differential equations coupled with the TOV equations,
\begin{eqnarray}
  \frac{dH}{dr} &=& \beta,\nonumber\\
  \frac{d\beta}{dr} &=& 2\left(1-2\frac{m_r}{r}\right)^{-1}H\nonumber\\
  & &{} \times \Big\{-2\pi[5\epsilon+9P+f(\epsilon+P)]\nonumber\\
  &&{} + \frac{3}{r^2}+2\left(1-2\frac{m_r}{r}\right)^{-1}\left(\frac{m_r}{r^2}+4\pi rP\right)^2\Big\}\nonumber\\
  &&{} + \frac{2\beta}{r}\left(1-2\frac{m_r}{r}\right)^{-1}\left\{\frac{m_r}{r}+2\pi r^2(\epsilon-P)-1\right\},\,\nonumber\\
  \label{HbetaEq}
\end{eqnarray}
where $f={\rm d}\epsilon/{\rm d}P$.

\begin{figure}
\centering
\subfigure[]{
\centering
\label{Fig:8a}
\includegraphics[width=0.45\textwidth]{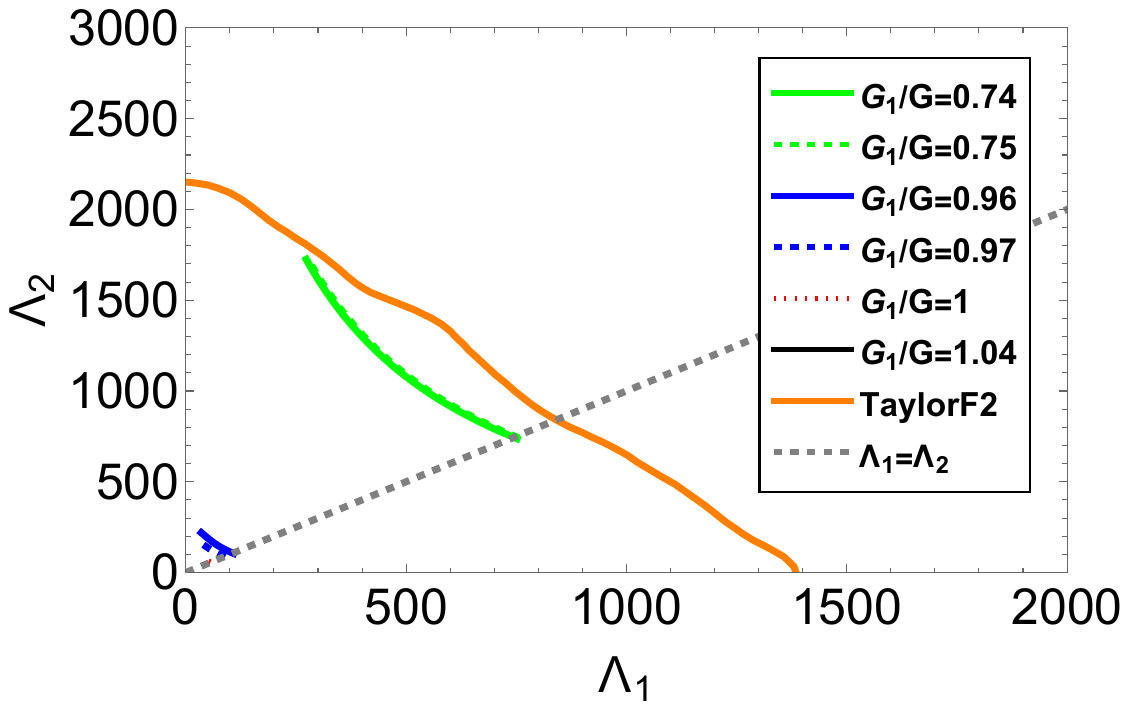}
}
\subfigure[]{
\centering
\label{Fig:8b}
\includegraphics[width=0.45\textwidth]{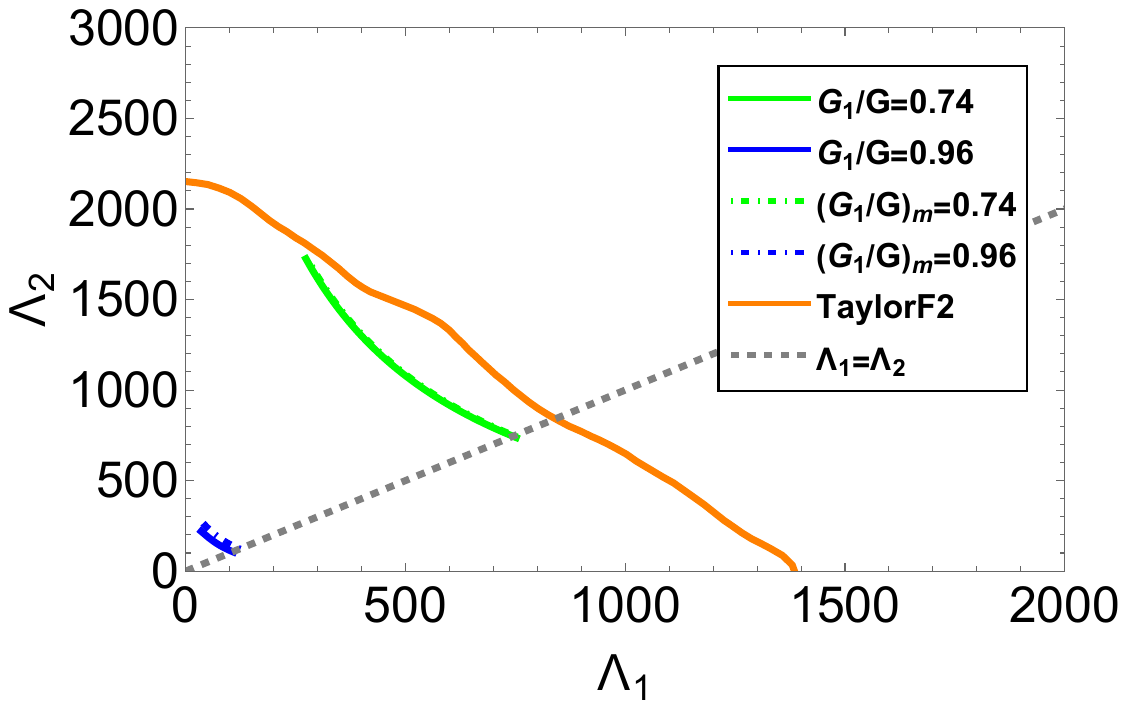}
}
\subfigure[]{
\centering
\label{Fig:8c}
\includegraphics[width=0.45\textwidth]{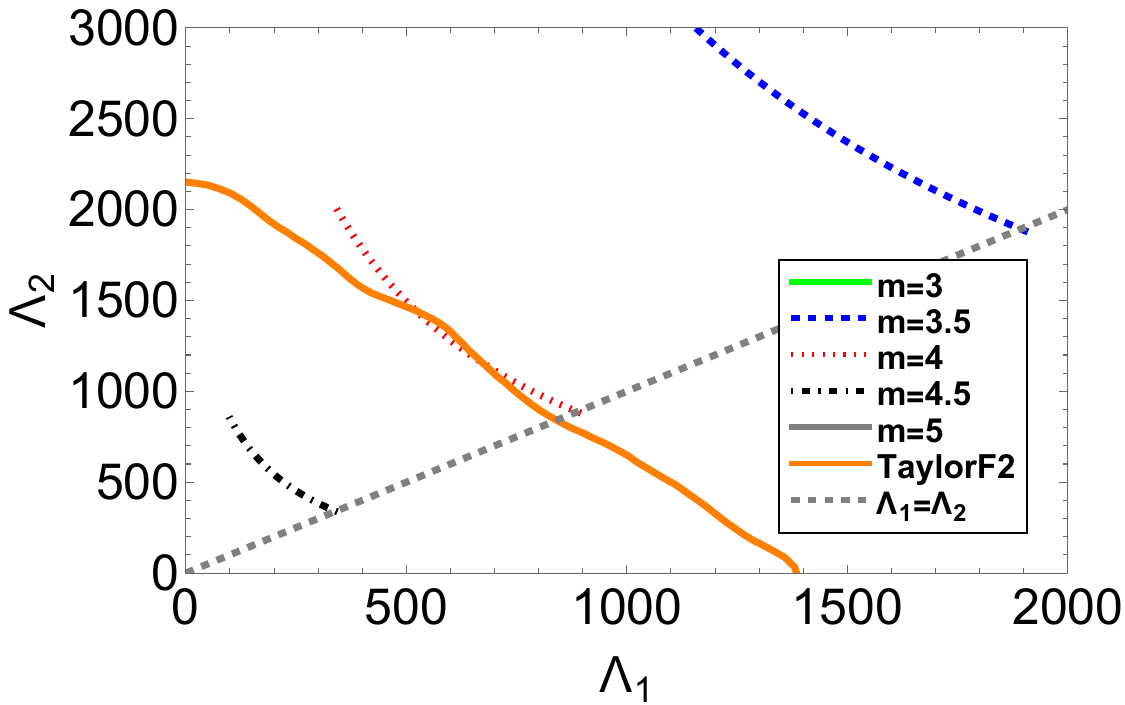}
}
\caption{The $\Lambda_1-\Lambda_2$ based on the EOSs in this work. The line styles correspond to the same scenarios as those in Fig.~\ref{Fig:7}. The orange line refers to the constraint from the waveform model of TaylorF2~\cite{PhysRevX.9.011001}.}
\label{Fig:8}
\end{figure}

The $\Lambda_1-\Lambda_2$ relation for the binary components in the GW170817 event during the inspiral phase based on our EOSs are shown in
Fig.~\ref{Fig:8}, where each curve corresponds to the same scenario as that in Fig.~\ref{Fig:7}, respectively. The orange curve represents the constraint derived from the waveform model TaylorF2 in GW170817~\cite{PhysRevX.9.011001}. If the $\Lambda_1-\Lambda_2$ relation falls to
the bottom-left of the orange curve, it can satisfy the
constraint from GW170817. In Fig.~\ref{Fig:8a}, all $\Lambda_1-\Lambda_2$ relations with $G_1/G = 0.74, 0.75, 0.96, 0.97, 1$ satisfy the constraint from GW170817. Notably, for $0.74\leq G_1/G\leq0.75$, the $\Lambda_1-\Lambda_2$ relations are very close to the
constraint boundary, whereas for $G_1/G\geq0.96$, they are far from it. In Fig.~\ref{Fig:8b}, we find that the VP has a negligible influence on the results. In Fig.~\ref{Fig:8c}, the $\Lambda_1-\Lambda_2$ curves shift towards the bottom-left as the current quark mass increases. However, even for $m = 4$ MeV, the curve still fails to satisfy the constraint from GW170817, implying that the constraint can only be satisfied for $m>4$ MeV. It is worth noting that for $m = 3$ MeV, the $\Lambda_1-\Lambda_2$ curve is located
in the upper-right region of the plot, far exceeding the
constraint boundary. Conversely, for $m = 5$ MeV, the
maximum star mass is $1.34 M_{\odot}$, and the corresponding $\Lambda_1-\Lambda_2$ relation cannot be calculated in this case, because simulations of GW170817 indicate that at least one compact star in the binary should possess a mass larger than $1.36 M_{\odot}$. Therefore, the curves corresponding to $m = 3$, 5 MeV are not shown in Fig.~\ref{Fig:8c}. The detailed information of parameter sets and properties of non-strange quark stars are shown in Table~\ref{parasetandnsqs}.

\begin{widetext}
\begin{center}
\begin{table}
\caption{The parameter sets and properties of corresponding non-strange quark stars: current quark mass $m$, $u$ quark condensate $-\langle\bar{u}u\rangle^{1/3}$, ultraviolet momentum cutoff $\Lambda_{\rm UV}$, coupling constant $G$, the ratio $G_1/G$, weight parameter $G_2$, $B_0^{1/4}$ calculated with the strict Winger solution, $(B_0^{1/4})^\prime$ calculated with the approximate Winger solution, maximum gravitational mass $M_{\rm m}$, radius $R_m$, radius of 1.4$M_{\odot}$ star $R(1.4)$, dimensionless tidal deformability of 1.4$M_{\odot}$ star $\Lambda(1.4)$ calculated with the strict Winger solution, and the corresponding quantities calculated with the approximate Winger solution, including $M_{\rm m}^\prime$, $R_m^\prime$, $R^\prime(1.4)$, and $\Lambda^\prime(1.4)$.}\label{parasetandnsqs}
\begin{tabular}{p{0.9cm} p{1.2cm} p{0.9cm} p{1.1cm} p{1.0cm}p{1.2cm}p{0.9cm}p{1.2cm}p{0.8cm}p{0.8cm}p{1.0cm}p{1.0cm}p{0.8cm}p{0.7cm}p{1.1cm}p{1.0cm}}
\hline\hline
$\quad m$&$-\langle\bar{u}u\rangle^{1/3}$&$\,\,\Lambda_{\rm UV}$&$\quad\,\,\, G$&$\,\,\, G_1/G$&$\quad\,\,\, G_2$&$\,\,\,B_0^{1/4}$&$\,\,(B_0^{1/4})^\prime$&$\,M_{m}$&$\,R_m$&$R(1.4)$&$\Lambda(1.4)$&$\,M_{m}^\prime$&$\,R_m^\prime$&$R^\prime(1.4)$&$\Lambda^\prime(1.4)$\\
$[{\rm MeV}]$&$\,\,\,[{\rm MeV}]$&$[{\rm MeV}]$&$[{\rm GeV}^{-2}]$&$\quad[-]$&$[{\rm GeV}^{-5}]$&$\,\,[{\rm MeV}]$&$\,\,\,[{\rm MeV}]$&$[M_{\odot}]$&$[{\rm km}]$&$\,\,[{\rm km}]$&$\,\,\,[-]$&$[M_{\odot}]$&$[{\rm km}]$&$\,\,\,[{\rm km}]$&$\,\,\,\,[-]$\\
\hline
\multirow{6}{*}{$\,\,\,\,4.1$}&\multirow{6}{*}{$\quad267$}&\multirow{6}{*}{$\,\,758$}&\multirow{6}{*}{$\quad3.31$}&$\,\,\,\,\,0.74$&$\,\,\,-22.7$&$\,\,40.4$&$\,\,\,\,62.9$&1.95&11.32&$\,\,11.58$&$\,639.8$&1.97&11.35&$\,\,\,11.57$&$\,\,642.6$\\
                                                                                                               &&&&$\,\,\,\,\,0.75$&$\,\,\,-21.8$&$\,\,63.7$&$\,\,\,\,75.0$&1.97&11.35&$\,\,11.58$&$\,644.0$&1.98&11.38&$\,\,\,11.57$&$\,\,646.8$\\
                                                                                                               &&&&$\,\,\,\,\,0.96$&$\,\,\,-3.5$&$\,\,139.7$&$\,\,\,\,138.3$&1.52&8.60&$\,\,8.78$&$\,85.7$&1.55&8.73&$\,\,\,8.93$&$\,\,99.3$\\
                                                                                                               &&&&$\,\,\,\,\,0.97$&$\,\,\,-2.6$&$\,\,140.9$&$\,\,\,\,139.8$&1.47&8.43&$\,\,8.56$&$\,69.0$&1.51&8.56&$\,\,\,8.73$&$\,\,81.8$\\
                                                                                                               &&&&$\quad\,\,\,1$&$\quad\,\,\,0$&$\,\,144.7$&$\,\,\,\,144.0$&1.38&8.07&$\,\,\,\,\,-$&$\,\,\,\,-$&1.38&8.07&$\quad-$&$\quad-$\\
                    \hline
                $\,\,\,\,3.0$&$\quad296$&$\,\,928$&$\quad2.07$&$\,\,\,\,\,0.84$&$\,\,\,-6.4$&$\,\,47.3$&$\,\,\,\,70.1$&2.71&15.66&$\,\,14.83$&$\,3290$&2.76&15.76&$\,\,\,14.77$&$\,\,3255$\\
                    \hline
                $\,\,\,\,3.5$&$\quad281$&$\,\,842$&$\quad2.59$&$\,\,\,\,\,0.80$&$\,\,\,-11.6$&$\,\,48.8$&$\,\,\,\,68.6$&2.35&13.58&$\,\,13.32$&$\,1655$&2.38&13.63&$\,\,\,13.30$&$\,\,1651$\\
                    \hline
                $\,\,\,\,4.0$&$\quad269$&$\,\,771$&$\quad3.18$&$\,\,\,\,\,0.76$&$\,\,\,-19.6$&$\,\,62.9$&$\,\,\,\,74.8$&2.03&11.71&$\,\,11.87$&$\,766.5$&2.05&11.74&$\,\,\,11.86$&$\,\,769.1$\\
                    \hline
                $\,\,\,\,4.5$&$\quad259$&$\,\,711$&$\quad3.90$&$\,\,\,\,\,0.71$&$\,\,\,-32.7$&$\,\,72.0$&$\,\,\,\,79.9$&1.72&9.94&$\,\,10.34$&$\,285.7$&1.73&9.95&$\,\,\,10.34$&$\,\,288.3$\\
                    \hline
                $\,\,\,\,5.0$&$\quad250$&$\,\,657$&$\quad4.82$&$\,\,\,\,\,0.65$&$\,\,\,-54.3$&$\,\,85.5$&$\,\,\,\,90.0$&1.34&8.28&$\quad-$&$\,\,\,\,\,-$&1.35&8.29&$\quad\,\,-$&$\quad-$\\
\hline\hline
\end{tabular}
\end{table}
\end{center}
\end{widetext}

\begin{figure}
\includegraphics[width=0.47\textwidth]{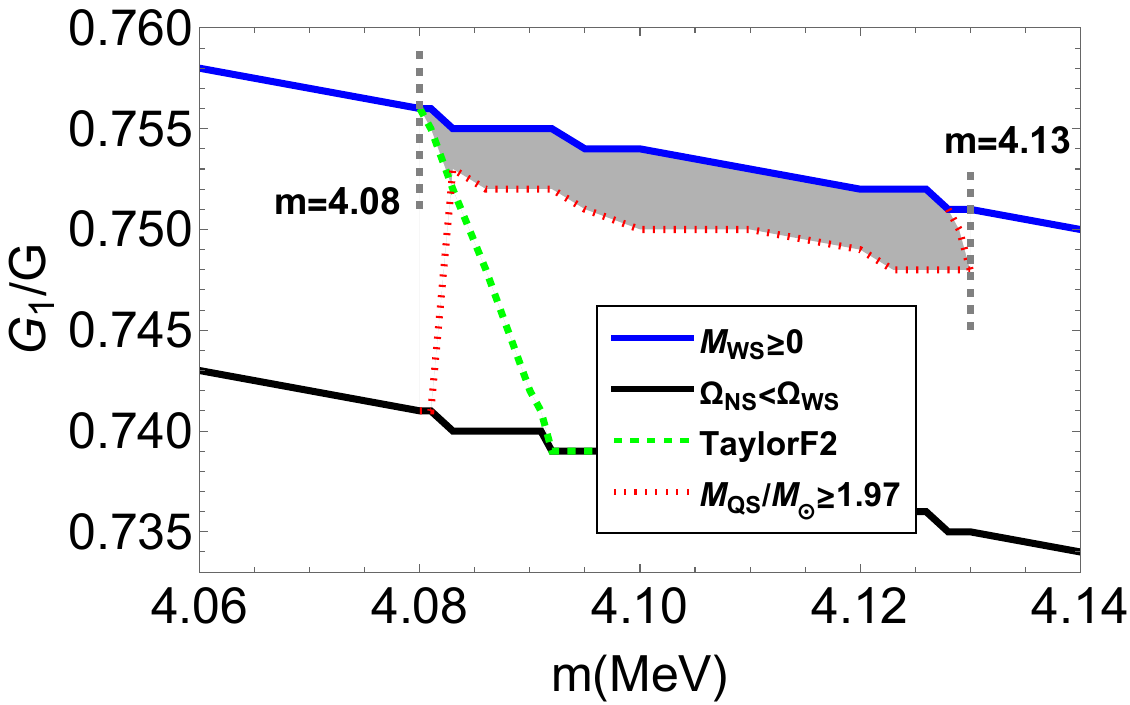}
\caption{The parameter space of $G_1/G - m$ restricted
by the four constraints.}
\label{Fig:9}
\end{figure}

To investigate the possibility of the existence of nonstrange quark stars and to determine the proportion of the contribution from quark condensate to the gluon
propagator, we employ four constraints derived from theory and compact star observations to restrict the model parameter space, and the result is shown as the gray region in Fig.~\ref{Fig:9}.
The four constraints are: (i) The quark gap equation possesses a non-negative Winger solution at $T=\mu=0$. (ii) The thermodynamic potential of the Nambu solution is lower than that of the Winger solution at $T=\mu=0$. (iii) Constraints on the tidal deformability of the binary system in GW170817 based on the TaylorF2 waveform model~\cite{PhysRevX.9.011001}. (iv) Constraints on the maximum pulsar mass ($1.97 M_{\odot}$) from PSR J0348+0432~\cite{Antoniadis1233232}. Satisfying the first two theoretical constraints ensures the stable existence of the Nambu solution, which corresponds to the physical reality. On the other hand, the VP can also be calculated if the first two constraints are satisfied.

In Fig.~\ref{Fig:9}, it can be seen that the model parameter space is restricted to a small region with $4.08 \leq m \leq 4.13$ and $0.748\leq G_1/G\leq0.756$. Consequently, the proportion of the contribution from quark condensate to the gluon propagator is
$G_2\langle\bar{\psi}\psi\rangle/G=1-G_1/G\sim0.25$. Our result indicates that the non-strange quark stars can exist in the universe. Furthermore, given that $G_1/G \sim 0.75$, it can be inferred from Fig.~\ref{Fig:3b} that the chiral phase transition occurring in the interior of massive
non-strange quark stars is a first-order phase transition rather than a smooth crossover.

\section{Summary and discussion}\label{four}
In this study, inspired by the hypothesis that quark matter could be non-strange, we employ a modified two-flavor NJL model to investigate the EOS of non-strange quark stars, with a particular focus on the influence of VP.
In the modified NJL model, both the thermodynamic potential and the four-fermion interaction coupling strength $G$ are modified to satisfy thermodynamic consistency and the essential requirements of QCD.

Our study indicates that the VP calculated with the modified NJL model for a small $G_1/G$ ratio ($0.74-0.75$) differs significantly from the result obtained with the conventional NJL model, which corresponds to $ G_1/G=1$ in the modified version. Consequently, the properties of the chiral phase transition, the EOS, the sound velocity, and the $M-R$ relation of non-strange quark stars differ substantially between these two scenarios, demonstrating the impact of this modification. Specifically, when $G_1/G$ is small, the sound velocity at low densities is significantly higher than that in the normal NJL model, and a first-order chiral phase transition rather than a smooth crossover occurs.

In the case of small $G_1/G$, the ordering of two bag constants $B_0^{1/4}(M_W)$ and $B_0^{1/4}(m)$ is reversed compared to the case of large $G_1/G$, with a clear gap in the numerical ranges between these two cases.
It is worth noting that many previous studies found a positive correlation between the bag constant and the stiffness of the EOS, that is, a smaller bag constant yielded a stiffer EOS and consequently a larger maximum mass for compact stars~\cite{PhysRevD.101.063023,PhysRevD.100.043018,PhysRevD.103.063018,PhysRevD.101.043003}. We also investigate the influence of the bag constant on the EOS and quark stars, showing that if $G_1/G$ is small when a first-order chiral phase transition happens, an increase in the bag constant will stiffen the EOS at low energy densities, while softening it at high energy densities. Given that a larger bag constant results in a larger maximum mass of compact stars in this case, which contradicts previous conclusions, we argue that the influence of the bag constant (or VP) on the EOS at low energy densities plays a dominant role.

To determine the proportion of the contribution from quark condensate to the gluon propagator and to investigate the possibility of the existence of non-strange quark stars, this study employs four constraints obtained from theoretical analysis and pulsar observations to restrict the model parameter space. It is found that non-strange quark stars can exist when $4.08\leq m\leq4.13$ MeV and the contribution from quark condensate to the gluon propagator accounts for approximately 25\%. In contrast to earlier findings that constrained the bag constant to $140<B_0^{1/4}<145$ MeV via the linear sigma model~\cite{PhysRevD.102.083003}, our study restricts it to $64<B_0^{1/4}<69$ MeV.

The maximum mass of non-strange quark stars is found to be $1.98 M_{\odot}$, and a first-order chiral phase transition will happen in massive stars. Furthermore, our study suggests that the merging compact binary in the GW170817 event consists of non-strange quark stars, and the tidal deformability, compared to the previous constraint of $\Lambda(1.4)\leq800$~\cite{PhysRevLett.119.161101}, is further constrained to $\Lambda(1.4)\leq646$. The pulsars with masses exceeding $1.98 M_{\odot}$ discovered in recent years, such as PSR J0740+6620, PSR J2215+5135, and PSR J0952-0607, might not be non-strange quark stars according to this study. In the future, more astronomical observations regarding pulsar masses, radii, and tidal deformability are expected to help us further determine the composition of compact stars.

Finally, we want to say that the present study modified the four-fermion interaction coupling strength as a function of quark condensate. It is also interesting to investigate the gluon condensate effect. This can be accomplished by using, for example, the dilaton compensator approach~\cite{Sheng:2023rnd} in the sense of Brown-Rho scaling~\cite{Brown:1991kk}. We will come to this issue in the next publication.

\acknowledgments

This work is supported in part by the
National Key Program for Science and Technology Research
Development (2023YFB3002500), the National Natural Science Foundation of China (under Grants No. 12005192, No. 12547104, and No. 12233002), the Natural Science Foundation of Henan Province of China (No. 242300421375, No. 262300421827, and No. 262300421877), the Project funded by China Postdoctoral Science Foundation (No. 2020M672255, No. 2020TQ0287), the Gusu Talent Innovation Program under Grant No. ZXL2024363. Y.F.H also acknowledges the support from the Xinjiang Tianchi Program.

\bibliography{reference}
\end{document}